\documentclass[aps,a4paper,superscriptaddress,showpacs,preprintnumbers,amsmath,amssymb]{revtex4}

\usepackage{ulem}

\usepackage{psfrag} \usepackage{graphicx} \usepackage{dcolumn}
\usepackage{color} \usepackage{latexsym,amsfonts} \usepackage{bm}
\usepackage{amssymb}
\begin{document}

\title{Gauge and Infrared Properties of Hadronic Structure of
  Nucleon\\ in Neutron Beta Decay to Order $O(\alpha/\pi)$ in Standard
  $V - A$ Effective Theory\\ with QED and Linear Sigma Model of Strong
  Low--Energy Interactions}

\author{A. N. Ivanov}\email{ivanov@kph.tuwien.ac.at}
\affiliation{Atominstitut, Technische Universit\"at Wien, Stadionallee
  2, A-1020 Wien, Austria}
\author{R.~H\"ollwieser}\email{roman.hoellwieser@gmail.com}
\affiliation{Atominstitut, Technische Universit\"at Wien, Stadionallee
  2, A-1020 Wien, Austria}\affiliation{Department of Physics,
  Bergische Universit\"at Wuppertal, Gaussstr. 20, D-42119 Wuppertal,
  Germany} \author{N. I. Troitskaya}\email{natroitskaya@yandex.ru}
\affiliation{Atominstitut, Technische Universit\"at Wien, Stadionallee
  2, A-1020 Wien, Austria}
\author{M. Wellenzohn}\email{max.wellenzohn@gmail.com}
\affiliation{Atominstitut, Technische Universit\"at Wien, Stadionallee
  2, A-1020 Wien, Austria} \affiliation{FH Campus Wien, University of
  Applied Sciences, Favoritenstra\ss e 226, 1100 Wien, Austria}
\author{Ya. A. Berdnikov}\email{berdnikov@spbstu.ru}\affiliation{Peter
  the Great St. Petersburg Polytechnic University, Polytechnicheskaya
  29, 195251, Russian Federation}

\date{\today}

\begin{abstract}
Within the standard $V - A$ theory of weak interactions, Quantum
Electrodynamics (QED) and the linear $\sigma$--model (L$\sigma$M) of
strong low--energy hadronic interactions we analyse gauge and infrared
properties of hadronic structure of the neutron and proton in the
neutron $\beta^-$--decay to leading order in the large nucleon mass
expansion. We show that the complete set of Feynman diagrams
describing radiative corrections of order $O(\alpha/\pi)$, induced by
hadronic structure of the nucleon, to the rate of the neutron
$\beta^-$--decay is gauge non--invariant and unrenormalisable. We show
that a gauge non--invariant contribution does not depend on the
electron energy in agreement with Sirlin's analysis of contributions
of strong low--energy interactions (Phys. Rev. {\bf 164}, 1767
(1967)). We show that infrared divergent and dependent on the electron
energy contributions from the neutron radiative $\beta^-$--decay and
neutron $\beta^-$--decay, caused by hadronic structure of the nucleon,
are cancelled in the neutron lifetime. Nevertheless, we find that
divergent contributions of virtual photon exchanges to the neutron
lifetime, induced by hadronic structure of the nucleon, are
unrenormalisable even formally. Such an unrenormalizability can be
explained by the fact that the effective $V - A$ vertex of
hadron--lepton current--current interactions is not a vertex of the
combined quantum field theory including QED and L$\sigma$M, which are
renormalizable theories. We assert that for a consistent gauge
invariant and renormalizable analysis of contributions of hadronic
structure of the nucleon to the radiative corrections of any order to
the neutron decays one has to use a gauge invariant and fully
renormalizable quantum field theory including the Standard Electroweak
Model (SEM) and the L$\sigma$M, where the effective $V - A$ vertex of
hadron--lepton current--current interactions is caused by the
$W^-$--electroweak--boson exchange.
\end{abstract}

\maketitle

\section{Introduction}
\label{sec:introduction}

Recently \cite{Ivanov2018b,Ivanov2018c} we have analysed gauge and
infrared properties of hadronic structure of the neutron and proton
in the neutron radiative $\beta^-$--decay (inner bremsstrahlung) to
order $O(\alpha/\pi)$, where $\alpha$ is the fine--structure constant
\cite{PDG2018}. Such an analysis we have carried out in the standard
$V - A$ effective theory of weak interactions
\cite{Feynman1958,Nambu1960} with Quantum Electrodynamics (QED) and
the linear $\sigma$--model (L$\sigma$M)
\cite{GellMann1960}--\cite{DeAlfaro1973}. We have shown
\cite{Ivanov2018b} that in the amplitude of the neutron
$\beta^-$--decay, calculated to one--hadron--loop approximation in the
infinite limit of the scalar $\sigma$--meson mass $m_{\sigma} \to
\infty$, to leading order in the large nucleon mass expansion and
after renormalization, the L$\sigma$M reproduces well--known Lorentz
structure of the matrix element of the hadronic $n\to p$ transition
(see also \cite{Leitner2006,Ivanov2018}), where strong low--energy
interactions are described by the axial coupling constant, the weak
magnetism \cite{Bilenky1959,Wilkinson1982} and the one--pion--pole
exchange \cite{Marshak1969}.

Our analysis of the amplitude of the neutron radiative
$\beta^-$--decay \cite{Ivanov2018b} has shown that to one--hadron loop
approximation the amplitude of the neutron radiative $\beta^-$--decay
is renormalizable and gauge invariant only to leading order in the
large nucleon mass expansion. However, to next--to--leading order in
the large nucleon mass expansion there are some one--hadron--loop
contributions, which violate both gauge invariance and
renormalizability.

In turn, in \cite{Ivanov2018c} keeping only the leading order
contributions in the large nucleon mass expansion to the amplitude of
the neutron radiative $\beta^-$--decay we have found an infrared
divergent contribution to the neutron lifetime, caused by hadronic
structure of the nucleon through the one--pion--pole exchange. We have
found that the order of this infrared divergence relative to Sirlin's
infrared divergent contribution \cite{Sirlin1967,Sirlin1975a} is of
about $10^{-5}$. This confirms validity and high confidence level of
the contribution of hadronic structure of the nucleon to the radiative
corrections of the neutron $\beta^-$--decay, calculated by Sirlin
\cite{Sirlin1967,Sirlin1978} to leading order in the large nucleon
mass expansion. The infrared divergent contribution of hadronic
structure of the nucleon, observed in \cite{Ivanov2018c}, arises the
problem of its necessary cancellation by the infrared divergent
contribution of virtual photon exchanges to the rate of the neutron
$\beta^-$--decay. Following \cite{Ivanov2018b} we may have doubts that
such a cancellation can be observed in the standard $V - A$ effective
theory of weak interactions with QED and the L$\sigma$M. This is
because the vertex of the effective $V - A$ hadron--lepton
current--current interaction is not a vertex of the combined quantum
field theory including QED and the L$\sigma$M. According to
\cite{Ivanov2018b}, beforehand we may assume that such a cancellation
can be well observed in the combined quantum field theory including
the Standard Electroweak Model (SEM) \cite{PDG2018} and a
renormalizable theory of strong low--energy interactions (e.g. the
L$\sigma$M), where the effective vertex of $V - A$ hadron--lepton
current--current interaction is caused by the
$W^-$--electroweak--boson exchange. Nevertheless, this paper addresses
to the analysis of gauge and infrared properties of hadronic structure
of the nucleon in the radiative corrections of order $O(\alpha/\pi)$
to the neutron lifetime in the standard $V - A$ effective theory of
weak interactions with QED and the L$\sigma$M. Such an analysis is
required by the necessity to shed light on possibility of further
application of the standard $V - A$ effective theory of weak
interactions with QED and the L$\sigma$M to calculation of higher
order radiative corrections to the neutron decays \cite{Ivanov2017b}.

The paper is organized as follows. In section \ref{sec:betadecay} we
discuss the amplitude of the neutron $\beta^-$--decay and the
structure of the hadronic $n \to p$ transition, calculated in
\cite{Ivanov2018b} to one--hadron-loop approximation in the standard
$V - A$ effective theory of weak interactions and the L$\sigma$M. 
In section \ref{sec:strahlung} we outline the calculation of the
radiative corrections to order $O(\alpha/\pi)$ to the neutron lifetime
by Sirlin \cite{Sirlin1967} and calculate the contribution of hadronic
structure of the nucleon to the radiative corrections of order
$O(\alpha/\pi)$ to the neutron lifetime to leading order in the large
nucleon mass expansion and to order $1/m^2_{\pi}$, where $m_{\pi}$ is
the $\pi$--meson mass. We show that the contributions of the Feynman
diagrams, induced by hadronic structure of the nucleon, can be
decomposed into invariant and non--invariant parts under gauge
transformation of the photon propagator $D_{\alpha\beta}(k) \to
D_{\alpha\beta}(k) + c(k^2) k_{\alpha}k_{\beta}$, where $c(k^2)$ is an
arbitrary function of $k^2$ \cite{Sirlin1967}. We show that a gauge
non--invariant part of the Feynman diagrams is ultra--violet and
infrared divergent and does not depend on the electron energy. Such an
independence of the electron energy of a gauge non--invariant part of
the Feynman diagrams to order $O(\alpha/\pi)$, induced by hadronic
structure of the nucleon, agrees well with Sirlin's analysis of the
contributions of hadronic structure of the nucleon
\cite{Sirlin1967,Sirlin1978}.  We calculate the infrared divergent and
dependent on the electron energy contribution to the radiative
corrections, induced by hadronic structure of the nucleon and virtual
photon exchanges. We show that together with the infrared divergent
contribution of hadronic structure, coming from the neutron radiative
$\beta^-$--decay, the obtained expression for the neutron
$\beta^-$--decay leads to infrared independent contribution to the
neutron lifetime. However, because of the absence of any parameter,
which is able to absorb the divergent contributions independent of the
electron energy, the calculated radiative corrections to the neutron
lifetime, induced by hadronic structure of the nucleon and virtual
photon exchanges, are not renormalizable even formally and,
correspondingly, are not observable. In section \ref{sec:conclusion}
we discuss the obtained results and perspectives of the calculation of
renormalizable and observable radiative corrections to the neutron
lifetime, caused by hadronic structure of the nucleon, within a
combined quantum field theory including the Standard Electroweak Model
(SEM) and the L$\sigma$M.

\section{Amplitude of neutron $\beta^-$--decay in standard $V - A$ 
effective theory of weak interactions and L$\sigma$M of strong
low--energy interactions \cite{Ivanov2018b}}
\label{sec:betadecay}

The amplitude of the neutron $\beta^-$--decay takes the form
\cite{Ivanov2018b,Ivanov2018c}
\begin{eqnarray}\label{eq:1}
\hspace{-0.15in}M(n \to p e^-\bar{\nu}_e) = - G_V\langle
p(\vec{k}_p,\sigma_p)|J^+_{\mu}(0)|n(\vec{k}_n, \sigma_n)\rangle\,
\Big[\bar{u}_e\big(\vec{k}_e, \sigma_e\big) \gamma^{\mu}\big(1 -
  \gamma^5\big) v_{\nu}\big(\vec{k}_{\nu}, + \frac{1}{2}\big)\Big],
\end{eqnarray}
where $G_V = G_FV_{ud}/\sqrt{2}$ is the vector weak coupling constant,
and $G_F$ and $V_{ud}$ are the Fermi weak coupling constant and the
matrix element of the Cabibbo--Kobayashi--Maskawa (CKM) mixing matrix
\cite{PDG2018}, respectively, and $\bar{u}_e \gamma^{\mu}\big(1 -
\gamma^5\big) v_{\nu}$ is the matrix element of the leptonic $V - A$
current. Then, $\langle
p(\vec{k}_p,\sigma_p)|J^+_{\mu}(0)|n(\vec{k}_n, \sigma_n)\rangle$ is
the matrix element of the hadronic $n \to p$ transition.  Such a
matrix element, calculated in the limit $m_{\sigma} \to \infty$, to
leading order in the large nucleon mass expansion and after
renormalization, has the following Lorentz structure
\cite{Ivanov2018b}
\begin{eqnarray}\label{eq:2}
\langle p(\vec{k}_p,\sigma_p)|J^+_{\mu}(0)|n(\vec{k}_n,
\sigma_n)\rangle = \bar{u}_p\big(\vec{k}_p,
\sigma_p\big)\Big(\gamma_{\mu}\big(1 - g_A\gamma^5\big) +
\frac{\kappa}{2 m_N}\, i \sigma_{\mu\nu}q^{\nu} - \frac{2\,g_A m_N
}{m^2_{\pi} - q^2}\, q_{\mu} \gamma^5 \Big)\,u_n\big(\vec{k}_n,
\sigma_n\big),
\end{eqnarray}
where $\bar{u}_p$ and $u_n$ are the Dirac wave functions of the free
proton and neutron. The contributions of strong low--energy
interactions in Eq.(\ref{eq:2}) are presented by the axial coupling
constant $g_A$, the isovector anomalous magnetic moment of the nucleon
$\kappa$ and the one--pion--pole exchange.  The contribution of the
one--pion--pole exchange is natural in the standard $V - A$ effective
theory of weak interactions with L$\sigma$M, describing strong
low--energy interactions. Such a contribution is also required by
local conservation of the axial--vector hadronic current in the limit
$m_{\pi} \to 0$ \cite{Feynman1958,Nambu1960}
\begin{eqnarray}\label{eq:3}
\lim_{m_{\pi} \to\, 0}q^{\mu} \langle
p(\vec{k}_p,\sigma_p)|J^+_{\mu}(0)|n(\vec{k}_n, \sigma_n)\rangle =
\bar{u}_p\big(\vec{k}_p, \sigma_p\big)\big(- 2 m_N g_A +
2m_N g_A\big)\gamma^5\,u_n\big(\vec{k}_n, \sigma_n\big) = 0,
\end{eqnarray}
where the first term in brackets comes from the nucleon axial--vector
current by taking into account the Dirac equations for the free proton
and neutron, whereas the second term is caused by the one--pion--pole
exchange.  The one--pion--pole exchange contribution appears also in
the current algebra approach \cite{Marshak1969} (see also
\cite{Adler1968}).  The term with the Lorentz structure
$i\sigma_{\mu\nu}q^{\nu}/2m_N$, where $m_N$ is the nucleon mass,
describes the contribution of the weak magnetism
\cite{Bilenky1959,Wilkinson1982} with the isovector anomalous magnetic
moment of the nucleon $\kappa$.
\begin{figure}
\centering \includegraphics[height=0.12\textheight]{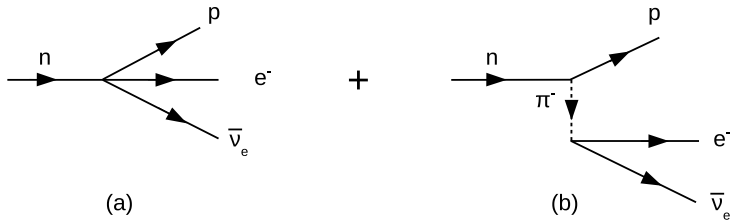}
  \caption{The Feynman diagrams, describing in the standard $V - A$
    effective theory of weak interactions the amplitude of the neutron
    $\beta^-$--decay, defined in the limit $m_{\sigma} \to \infty$, to
    leading order in the large nucleon mass expansion and after
    renormalization in the L$\sigma$M.}
\label{fig:fig1}
\end{figure} 
The amplitude of the neutron $\beta^-$--decay Eq.(\ref{eq:1}) with the
matrix element of the hadronic $n \to p$ transition, given by
Eq.(\ref{eq:2}), can be formally represented by the Feynman diagrams in
Fig.\,\ref{fig:fig1}. The first two terms in Eq.(\ref{eq:2}) are
described by the Feynman diagram in Fig.\,\ref{fig:fig1}a, whereas the
third term is given by the Feynman diagram in Fig.\,\ref{fig:fig1}b.

\section{Radiative corrections to amplitude of neutron $\beta^-$--decay,
defined by virtual photon exchanges between charged particles in
Fig.\,\ref{fig:fig1}}
\label{sec:strahlung}

\begin{figure}
\centering \includegraphics[height=0.45\textheight]{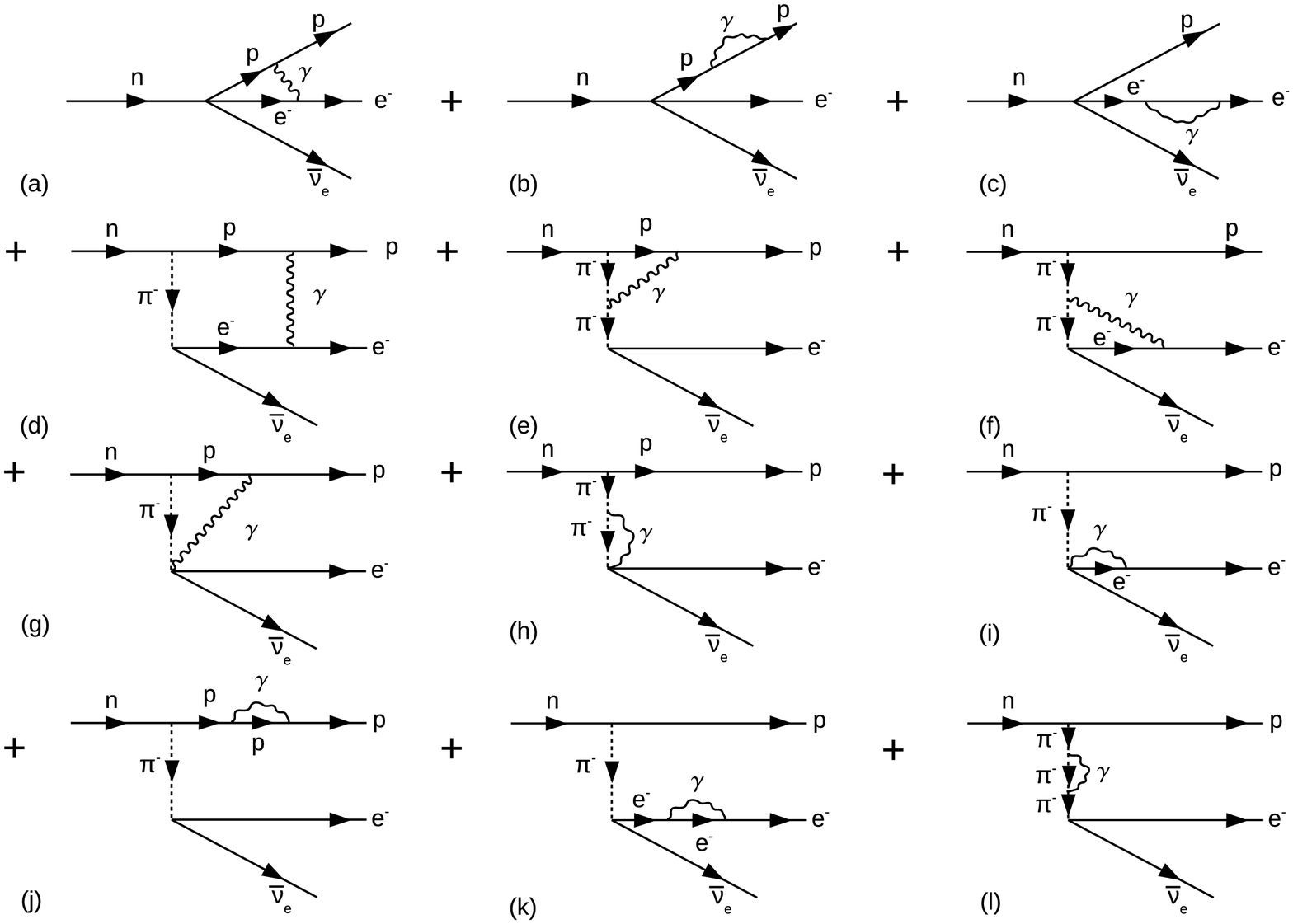}
\centering \includegraphics[height=0.115\textheight]{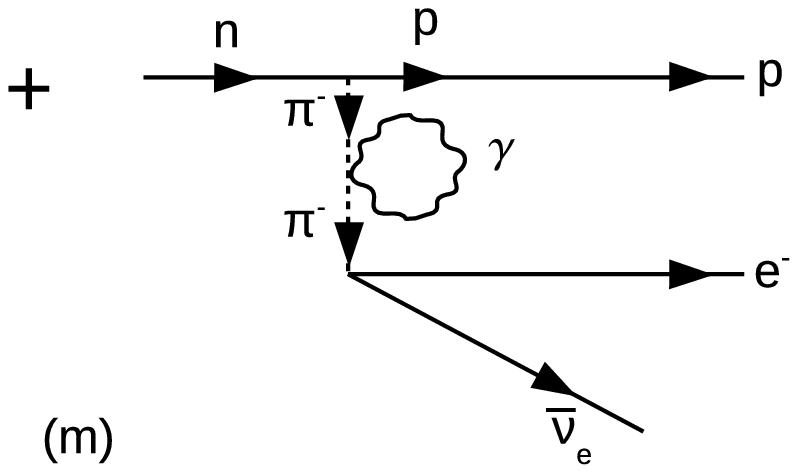}
  \caption{The Feynman diagrams, describing in the standard $V - A$
    effective theory of weak interactions the amplitude of the neutron
    $\beta^-$--decay, defined in the limit $m_{\sigma} \to \infty$, to
    leading order in the large nucleon mass expansion and after
    renormalization in the L$\sigma$M and QED.}
\label{fig:fig2}
\end{figure}

Inserting virtual photon exchanges between charged particles in the
Feynman diagrams in Fig.\,\ref{fig:fig1} we arrive at the set of
Feynman diagrams in Fig.\,\ref{fig:fig2}. Formally such a set of
Feynman diagrams can be obtained from the Feynman diagrams in Fig.\,5
in Ref.\cite{Ivanov2018b} with photon lines hooked by lines of charged
particles. The Feynman diagrams in Fig.\,\ref{fig:fig2}g -
Fig.\,\ref{fig:fig2}i are caused by the axial--vector hadronic current
$-ie\,f_{\pi}\pi^(x)A_{\mu}(x)$, induced in the phase of spontaneously
broken chiral $SU(2)\times SU(2)$ symmetry, where $f_{\pi}$,
$\pi^-(x)$ and $A_{\mu}(x)$ are the leptonic pion--coupling constant
and the field operators of the $\pi^-$--meson and electromagnetic
field, respectively \cite{Ivanov2018b}.
Following Sirlin \cite{Sirlin1967} for the calculation of the Feynman
diagram in Fig.\,\ref{fig:fig2}a, Fig.\,\ref{fig:fig2}b and
Fig.\,\ref{fig:fig2}c we neglect the contribution of the weak
magnetism. The contribution of the Feynman diagrams in
Fig.\,\ref{fig:fig2}a - Fig.\,\ref{fig:fig2}c has been calculated by
Sirlin \cite{Sirlin1967}. For details of this calculation we refer to
Appendices C - F of Ref.\cite{Ivanov2013}. The contribution of the
Feynman diagrams in Fig.\,\ref{fig:fig2}a - Fig.\,\ref{fig:fig2}c is
equal to \cite{Ivanov2013}
\begin{eqnarray}\label{eq:4}
\hspace{-0.3in}&&M(n \to p e^-\bar{\nu}_e)_{\rm Fig.\,\ref{fig:fig2}a
  - Fig.\,\ref{fig:fig2}c} = - 2m_N\,G_V\Big(1 +
\frac{\alpha}{2\pi}\,d_V\Big)\Big\{\Big(1 +
\frac{\alpha}{2\pi}\,f_{\beta}(E_e,\mu)\Big) [\varphi^{\dagger}_p
  \varphi_n][\bar{u}_e\,\gamma^0(1 -
  \gamma^5)v_{\bar{\nu}}]\nonumber\\
 \hspace{-0.3in}&& + g_A\Big(1 + \frac{\alpha}{2\pi}\,d_A\Big) \Big(1
 + \frac{\alpha}{2\pi}\,f_{\beta}(E_e,\mu) \Big)[\varphi^{\dagger}_p
   \vec{\sigma}\,\varphi_n]\cdot [\bar{u}_e \vec{\gamma}\,(1 -
   \gamma^5)v_{\bar{\nu}}] -
 \frac{\alpha}{2\pi}\,g_F(E_e)\,[\varphi^{\dagger}_p
   \varphi_n][\bar{u}_e\,(1 - \gamma^5)v_{\bar{\nu}}]\nonumber\\
\hspace{-0.3in}&&- \frac{\alpha}{2\pi}\,g_A\,
g_F(E_e)[\varphi^{\dagger}_p \vec{\sigma}\,\varphi_n]\cdot [\bar{u}_e
  \gamma^0\vec{\gamma}\,(1 - \gamma^5)v_{\bar{\nu}}] \Big\},
\end{eqnarray}
where $\varphi_p$ and $\varphi_n$ are the Pauli spinorial wave
functions of the proton and neutron, respectively. The functions
$f_{\beta}(E_e,\mu)$ and $g_F(E_e)$ are equal to \cite{Ivanov2013}
\begin{eqnarray}\label{eq:5}
f_{\beta}(E_e,\mu) &=& \frac{3}{2}{\ell
  n}\Big(\frac{m_p}{m_e}\Big) - \frac{11}{8} + 2 {\ell
  n}\Big(\frac{\mu}{ m_e}\Big)\Big[\frac{1}{2\beta}\,{\ell
    n}\Big(\frac{1 + \beta}{1 - \beta}\Big) - 1 \Big] -
\frac{1}{\beta}{\rm Li}_2\Big(\frac{2\beta}{1 + \beta}\Big) -
  \frac{1}{4\beta} {\ell n}^2\Big(\frac{1 + \beta}{1 - \beta}\Big) +
  \frac{1}{2\beta} {\ell n}\Big(\frac{1 + \beta}{1 -
    \beta}\Big),\nonumber\\
\hspace{-0.15in}g_F(E_e)&=& \frac{\sqrt{1 - \beta^2}}{2\beta}\, 
{\ell n}\Big(\frac{1 + \beta}{1 - \beta}\Big),
\end{eqnarray}
where $\beta$ is the electron velocity and ${\rm Li}_2(z)$ is the
Polylogarithmic function. The parameters $d_V$ and $d_A$ are
ultra--violet logarithmically divergent constants
\cite{Ivanov2013}. According to Sirlin \cite{Sirlin1967}, the
contribution of these constants should be absorbed by a formal
renormalization of the Fermi coupling constant $G_F$ and the axial
coupling constant $g_A$, respectively. However, as has been pointed
out in \cite{Ivanov2017b} we would prefer to cancel the contribution
of the constants $d_V$ and $d_A$ by the contribution of hadronic
structure of the neutron and proton, since such a cancellation is
required by gauge invariance of radiative corrections. The Feynman
diagrams in Fig.\, \ref{fig:fig2}a - Fig.\,\ref{fig:fig2}c define the
following contribution to the rate of the neutron $\beta^-$--decay
\begin{eqnarray}\label{eq:6}
\hspace{-0.3in}&&\lambda_{\beta}(E_0,\mu)_{\rm Fig.\,\ref{fig:fig2}a -
  Fig.\,\ref{fig:fig2}c} = (1 + 3
g^2_A)\,\frac{|G_V|^2}{\pi^3}\int^{E_0}_{m_e}dE_e \,\sqrt{E^2_e -
  m^2_e}\,E_e\,F(E_e, Z = 1)\,(E_0 - E_e )^2\Big(1+
\frac{\alpha}{\pi}\,\bar{g}_{\beta}(E_e,\mu)\Big),
\end{eqnarray}
where the function $g_{\beta}(E_e,\mu)$ is 
\begin{eqnarray}\label{eq:7}
\hspace{-0.15in}&&\bar{g}_{\beta}(E_e,\mu) = 2 {\ell
  n}\Big(\frac{\mu}{ m_e}\Big)\Big[\frac{1}{2\beta}\,{\ell
    n}\Big(\frac{1 + \beta}{1 - \beta}\Big) - 1 \Big] +
\frac{3}{2}{\ell n}\Big(\frac{m_p}{m_e}\Big) - \frac{11}{8} -
\frac{1}{\beta}{\rm Li}_2\Big(\frac{2\beta}{1 + \beta}\Big) -
\frac{1}{4\beta}\,{\ell n}^2\Big(\frac{1 + \beta}{1 - \beta}\Big) +
\frac{\beta}{2}\,{\ell n}\Big(\frac{1 + \beta}{1 -
  \beta}\Big).\nonumber\\
\hspace{-0.15in}&&
\end{eqnarray}
This result was obtained by Sirlin as a contribution of virtual photon
exchanges \cite{Sirlin1967} (for details see \cite{Ivanov2013}). Now
we may proceed to the calculation of the contributions of the Feynman
diagrams in Fig.\,\ref{fig:fig2}d - Fig.\,\ref{fig:fig2}m.  Since
after renormalization the contributions of the Feynman diagrams in
Fig.\,\ref{fig:fig2}j - Fig.\,\ref{fig:fig2}m to amplitude of the
neutron $\beta^-$--decay vanish \cite{Ivanov2017b}, we analyse the
Feynman diagrams in Fig.\,\ref{fig:fig2}d - Fig.\,\ref{fig:fig2}i
only. The analytical expressions of these diagrams are equal to
\begin{eqnarray}\label{eq:8}
\hspace{-0.3in}&&M(n \to p e^-\bar{\nu}_e)_{\rm Fig.\,\ref{fig:fig2}d}
= - G_V \,\Big\{ - 2 g_A m_N
e^2\Big[\bar{u}_p\big(\vec{k}_p,\sigma_p\big)\int
  \frac{d^4k}{(2\pi)^4i}\,\gamma^{\alpha}\frac{1}{m_N - \hat{k}_p -
    \hat{k} -
    i0}\,\gamma^5\,u_n\big(\vec{k}_n,\sigma_n\big)\Big]\nonumber\\
\hspace{-0.3in}&&\times\,\frac{(q + k)^{\mu}}{m^2_{\pi} - (q + k)^2 -
  i0}\, \Big[\bar{u}_e\big(\vec{k}_e,
  \sigma_e\big)\,\gamma^{\beta}\,\frac{1}{m_e - \hat{k}_e + \hat{k} -
    i0}\,\gamma^{\mu}(1 - \gamma^5)\, v_{\nu}\big(\vec{k}_{\nu}, +
  \frac{1}{2}\big)\Big]\,D_{\alpha\beta}(k),\nonumber\\
\hspace{-0.3in}&&M(n \to p e^-\bar{\nu}_e)_{\rm Fig.\,\ref{fig:fig2}e}
= - G_V \,\frac{2 g_A m_N\,e^2 q_{\mu}}{m^2_{\pi} - q^2 - i0}
\Big[\bar{u}_p\big(\vec{k}_p,\sigma_p\big) \int
  \frac{d^4k}{(2\pi)^4i}\,\gamma^{\alpha}\frac{1}{m_N - \hat{k}_p -
    \hat{k} -
    i0}\,\gamma^5\,u_n\big(\vec{k}_n,\sigma_n\big)\Big]\nonumber\\
\hspace{-0.3in}&&\times\, \frac{(2q + k)^{\beta}}{m^2_{\pi} - (q +
  k)^2 - i0}\, D_{\alpha\beta}(k)\, \Big[\bar{u}_e\big(\vec{k}_e,
  \sigma_e\big)\,\gamma^{\mu}(1 - \gamma^5)\,
  v_{\nu}\big(\vec{k}_{\nu}, + \frac{1}{2}\big)\Big],\nonumber\\
\hspace{-0.3in}&&M(n \to p e^-\bar{\nu}_e)_{\rm Fig.\,\ref{fig:fig2}f}
= - G_V \,\frac{ - 2 g_A m_N\,e^2}{m^2_{\pi} - q^2 - i0}
\Big[\bar{u}_p\big(\vec{k}_p,\sigma_p\big)\,\gamma^5\,
  u_n\big(\vec{k}_n,\sigma_n\big)\Big]\nonumber\\
\hspace{-0.3in}&&\times\,
\Big[\bar{u}_e\big(\vec{k}_e, \sigma_e\big)\int \frac{d^4k}{(2\pi)^4
    i}\gamma^{\alpha}\,\frac{1}{m_e - \hat{k}_e + \hat{k} -
    i0}\,\gamma^{\mu}(1 - \gamma^5)\, v_{\nu}\big(\vec{k}_{\nu}, +
  \frac{1}{2}\big)\Big]\,\frac{(2q + k)^{\beta} (q +
  k)_{\mu}}{m^2_{\pi} - (q + k)^2 - i0}\,D_{\alpha\beta}(k),\nonumber\\
\hspace{-0.3in}&&M(n \to p e^-\bar{\nu}_e)_{\rm Fig.\,\ref{fig:fig2}g}
= - 2 m_N\,G_V \,\Big\{2 g_A m_N e^2
\Big[\bar{u}_p\big(\vec{k}_p,\sigma_p\big) \int
  \frac{d^4k}{(2\pi)^4i}\,\gamma^{\alpha}\frac{1}{m_N - \hat{k}_p -
    \hat{k} -
    i0}\,\gamma^5\,u_n\big(\vec{k}_n,\sigma_n\big)\Big]\nonumber\\
\hspace{-0.3in}&&\times\, \frac{1}{m^2_{\pi} - (q + k)^2 - i0}\,
D_{\alpha\mu}(k)\, \Big[\bar{u}_e\big(\vec{k}_e,
  \sigma_e\big)\,\gamma^{\mu}(1 - \gamma^5)\,
  v_{\nu}\big(\vec{k}_{\nu}, + \frac{1}{2}\big)\Big]\Big\},\nonumber\\
\hspace{-0.3in}&&M(n \to p e^-\bar{\nu}_e)_{\rm Fig.\,\ref{fig:fig2}h}
= - G_V \,\frac{2 g_A m_N\,e^2}{m^2_{\pi} - q^2 - i0}
\Big[\bar{u}_p\big(\vec{k}_p,\sigma_p\big)\,\gamma^5\,
  u_n\big(\vec{k}_n,\sigma_n\big)\Big]\nonumber\\
\hspace{-0.3in}&&\times
  \Big[\bar{u}_e\big(\vec{k}_e, \sigma_e\big) \,\gamma^{\mu}(1 -
    \gamma^5)\, v_{\nu}\big(\vec{k}_{\nu}, + \frac{1}{2}\big)\Big]\int
  \frac{d^4k}{(2\pi)^4 i}\,\frac{(2q + k)^{\beta}}{m^2_{\pi} - (q +
    k)^2 - i0}\,D_{\mu\beta}(k),\nonumber\\
\hspace{-0.3in}&&M(n \to p e^-\bar{\nu}_e)_{\rm Fig.\,\ref{fig:fig2}i}
= - G_V \,\frac{ - 2 g_A m_N\,e^2}{m^2_{\pi} - q^2 - i0}
\Big[\bar{u}_p\big(\vec{k}_p,\sigma_p\big)\,\gamma^5\,
  u_n\big(\vec{k}_n,\sigma_n\big)\Big]\nonumber\\
\hspace{-0.3in}&&\times\, \Big[\bar{u}_e\big(\vec{k}_e,
  \sigma_e\big)\int \frac{d^4k}{(2\pi)^4
    i}\gamma^{\alpha}\,\frac{1}{m_e - \hat{k}_e + \hat{k} -
    i0}\,\gamma^{\mu}(1 - \gamma^5)\,D_{\alpha \mu}(k)\,
  v_{\nu}\big(\vec{k}_{\nu}, + \frac{1}{2}\big)\Big],
\end{eqnarray}
where $D_{\alpha\beta}(k)$ is the photon propagator \cite{Ivanov2013}
\begin{eqnarray}\label{eq:9}
D_{\alpha\beta}(k) = \frac{1}{k^2 + i0}\,\Big(\eta_{\alpha\beta} -
\xi\,\frac{k_{\alpha}k_{\beta}}{k^2}\Big).
\end{eqnarray}
Here $\eta_{\alpha\beta}$ is the Minkowski metric tensor and $\xi$ is
a gauge parameter. According to Sirlin \cite{Sirlin1967}, observable
radiative corrections to the neutron lifetime should be invariant
under a gauge transformation $D_{\alpha\beta}(k) \to
D_{\alpha\beta}(k) + c(k^2)\,k_{\alpha}k_{\beta}$, where $c(k^2)$ is
an arbitrary function of $k^2$. Following \cite{Ivanov2013} and
\cite{Ivanov2017b} we decompose the contributions of the Feynman
diagrams in Fig.\,\ref{fig:fig2}e - Fig.\,\ref{fig:fig2}i into
invariant and non--invariant parts with respect to a gauge
transformation $D_{\alpha\beta}(k) \to D_{\alpha\beta}(k) +
c(k^2)\,k_{\alpha}k_{\beta}$. We get
\begin{eqnarray*}
\hspace{-0.3in}&&M(n \to p e^-\bar{\nu}_e)^{(\rm gauge\,inv.)}_{\rm
  Fig.\,\ref{fig:fig2}e - Fig.\,\ref{fig:fig2}i} = - G_V\nonumber\\
\hspace{-0.3in}&&\times\,\Big\{ - 2 g_A m_N
e^2\Big[\bar{u}_p\big(\vec{k}_p,\sigma_p\big)\int
  \frac{d^4k}{(2\pi)^4i}\,\gamma^{\alpha}\frac{1}{m_N - \hat{k}_p -
    \hat{k} -
    i0}\,\gamma^5\,u_n\big(\vec{k}_n,\sigma_n\big)\Big]\nonumber\\
\hspace{-0.3in}&&\times\,\frac{(q + k)^{\mu}}{m^2_{\pi} - (q + k)^2 -
  i0}\, \Big[\bar{u}_e\big(\vec{k}_e,
  \sigma_e\big)\,\gamma^{\beta}\,\frac{1}{m_e - \hat{k}_e + \hat{k} -
    i0}\,\gamma^{\mu}(1 - \gamma^5)\, v_{\nu}\big(\vec{k}_{\nu}, +
  \frac{1}{2}\big)\Big]\,D_{\alpha\beta}(k)\nonumber\\
\hspace{-0.3in}&&+ \frac{2 g_A m_N\,e^2
  q^{\mu}}{m^2_{\pi} - q^2 - i0}
\Big[\bar{u}_p\big(\vec{k}_p,\sigma_p\big) \int
  \frac{d^4k}{(2\pi)^4i}\,\gamma^{\alpha}\frac{1}{m_N - \hat{k}_p -
    \hat{k} -
    i0}\,\gamma^5\,u_n\big(\vec{k}_n,\sigma_n\big)\Big]\nonumber\\
\hspace{-0.3in}&&\times\, \frac{(2q + k)^{\beta}}{m^2_{\pi} - (q +
  k)^2 - i0}\, D_{\alpha\beta}(k)\, \Big[\bar{u}_e\big(\vec{k}_e,
  \sigma_e\big)\,\gamma_{\mu}(1 - \gamma^5)\,
  v_{\nu}\big(\vec{k}_{\nu}, + \frac{1}{2}\big)\Big]\nonumber\\
\hspace{-0.3in}&& - \frac{2 g_A m_N\,e^2}{m^2_{\pi} - q^2 - i0}
\Big[\bar{u}_p\big(\vec{k}_p,\sigma_p\big)\,\gamma^5\,
  u_n\big(\vec{k}_n,\sigma_n\big)\Big]\nonumber\\
\hspace{-0.3in}&&\times\, \Big[\bar{u}_e\big(\vec{k}_e,
  \sigma_e\big)\int \frac{d^4k}{(2\pi)^4
    i}\gamma^{\alpha}\,\frac{1}{m_e - \hat{k}_e + \hat{k} -
    i0}\,\gamma_{\mu}(1 - \gamma^5)\, v_{\nu}\big(\vec{k}_{\nu}, +
  \frac{1}{2}\big)\Big]\,\frac{(2q + k)^{\beta} (q +
  k)^{\mu}}{m^2_{\pi} - (q + k)^2 - i0}\,D_{\alpha\beta}(k)\nonumber\\
\end{eqnarray*}
\begin{eqnarray}\label{eq:10}
\hspace{-0.3in}&&+ 2 g_A m_N e^2
\Big[\bar{u}_p\big(\vec{k}_p,\sigma_p\big) \int
  \frac{d^4k}{(2\pi)^4i}\,\gamma^{\alpha}\frac{1}{m_N - \hat{k}_p -
    \hat{k} -
    i0}\,\gamma^5\,u_n\big(\vec{k}_n,\sigma_n\big)\Big]\nonumber\\
\hspace{-0.3in}&&\times\, \frac{1}{m^2_{\pi} - (q + k)^2 - i0}\,
D_{\alpha\mu}(k)\, \Big[\bar{u}_e\big(\vec{k}_e,
  \sigma_e\big)\,\gamma^{\mu}(1 - \gamma^5)\,
  v_{\nu}\big(\vec{k}_{\nu}, + \frac{1}{2}\big)\Big]\nonumber\\
\hspace{-0.3in}&& + \frac{2 g_A m_N\,e^2}{m^2_{\pi} - q^2 - i0}
\Big[\bar{u}_p\big(\vec{k}_p,\sigma_p\big)\,\gamma^5\,
  u_n\big(\vec{k}_n,\sigma_n\big)\Big]\nonumber\\
\hspace{-0.3in}&&\times
  \Big[\bar{u}_e\big(\vec{k}_e, \sigma_e\big) \,\gamma^{\mu}(1 -
    \gamma^5)\, v_{\nu}\big(\vec{k}_{\nu}, + \frac{1}{2}\big)\Big]\int
  \frac{d^4k}{(2\pi)^4 i}\,\frac{(2q + k)^{\beta}}{m^2_{\pi} - (q +
    k)^2 - i0}\,D_{\mu\beta}(k)\nonumber\\
\hspace{-0.3in}&&- \frac{2 g_A m_N\,e^2}{m^2_{\pi} - q^2 - i0}
\Big[\bar{u}_p\big(\vec{k}_p,\sigma_p\big)\,\gamma^5\,
  u_n\big(\vec{k}_n,\sigma_n\big)\Big]\nonumber\\
\hspace{-0.3in}&&\times\, \Big[\bar{u}_e\big(\vec{k}_e,
  \sigma_e\big)\int \frac{d^4k}{(2\pi)^4
    i}\gamma^{\alpha}\,\frac{1}{m_e - \hat{k}_e + \hat{k} -
    i0}\,\gamma^{\mu}(1 - \gamma^5)\,D_{\alpha \mu}(k)\,
  v_{\nu}\big(\vec{k}_{\nu}, + \frac{1}{2}\big)\Big]\Big\}\nonumber\\
\hspace{-0.3in}&& + 2 g_A m_N e^2\,q^{\mu}
\Big[\bar{u}_p\big(\vec{k}_p,\sigma_p\big)\,\gamma^5\,
  u_n\big(\vec{k}_n,\sigma_n\big)\Big]\Big[\bar{u}_e\big(\vec{k}_e,
  \sigma_e\big)\,\gamma_{\mu}(1 - \gamma^5)\,
  v_{\nu}\big(\vec{k}_{\nu}, + \frac{1}{2}\big)\Big]\nonumber\\
\hspace{-0.3in}&&\times\,\Big(\int \frac{d^4k}{(2\pi)^4
  i}\,\frac{1}{m^2_{\pi} - (q + k)^2 - i0}\,\frac{1}{k^2 +
  i0}\,\eta^{\alpha\beta} D_{\alpha\beta}(k) - \frac{2}{m^2_{\pi} -
  q^2 - i0}\int \frac{d^4k}{(2\pi)^4 i}\,\frac{1}{k^2 +
  i0}\,\eta^{\alpha\beta} D_{\alpha\beta}(k)\Big)\Big\}
\end{eqnarray}
and 
\begin{eqnarray}\label{eq:11}
\hspace{-0.3in}&&M(n \to p e^-\bar{\nu}_e)^{(\rm
  gauge\,non-inv.)}_{\rm Fig.\,\ref{fig:fig2}e -
  Fig.\,\ref{fig:fig2}i} = - G_V \nonumber\\ 
\hspace{-0.3in}&&\times\,\Big\{- 2 g_A m_N e^2\,q^{\mu}
\Big[\bar{u}_p\big(\vec{k}_p,\sigma_p\big)\,\gamma^5\,
  u_n\big(\vec{k}_n,\sigma_n\big)\Big]\Big[\bar{u}_e\big(\vec{k}_e,
  \sigma_e\big)\,\gamma_{\mu}(1 - \gamma^5)\,
  v_{\nu}\big(\vec{k}_{\nu}, + \frac{1}{2}\big)\Big]\nonumber\\
\hspace{-0.3in}&&\times\,\Big(\int \frac{d^4k}{(2\pi)^4
  i}\,\frac{1}{m^2_{\pi} - (q + k)^2 - i0}\,\frac{1}{k^2 +
  i0}\,\eta^{\alpha\beta} D_{\alpha\beta}(k) - \frac{2}{m^2_{\pi} -
  q^2 - i0}\int \frac{d^4k}{(2\pi)^4 i}\,\frac{1}{k^2 +
  i0}\,\eta^{\alpha\beta} D_{\alpha\beta}(k)\Big)\Big\},
\end{eqnarray}
where $M(n \to p e^-\bar{\nu}_e)^{(\rm gauge\,inv.)}_{\rm
  Fig.\,\ref{fig:fig2}d - Fig.\,\ref{fig:fig2}f}$ and $M(n \to p
e^-\bar{\nu}_e)^{(\rm gauge\,non-inv.)}_{\rm Fig.\,\ref{fig:fig2}d -
  Fig.\,\ref{fig:fig2}f}$ are the amplitudes invariant and
non--invariant with respect to a gauge transformation
$D_{\alpha\beta}(k) \to D_{\alpha\beta}(k) +
c(k^2)\,k_{\alpha}k_{\beta}$, respectively.  Neglecting the
contributions of the terms of order $O(q^2/m^2_{\pi})$ as we have done
for the calculation of the infrared divergent contribution of hadronic
structure of the nucleon in the neutron radiative $\beta^-$--decay
\cite{Ivanov2018c}, we get
\begin{eqnarray}\label{eq:12}
\hspace{-0.3in}M(n \to p e^-\bar{\nu}_e)^{(\rm gauge\,non-inv.)}_{\rm
  Fig.\,\ref{fig:fig2}d - Fig.\,\ref{fig:fig2}f} &=& - G_V\,
\frac{2 g_A m_N m_e}{m^2_{\pi}}\,\frac{\alpha}{\pi}\,\Big\{ (4 - \xi)
\Big({\ell n}\frac{\Lambda}{m_N} + {\ell
  n}\frac{m_N}{\mu} - \frac{1}{4}\Big)\Big\}\nonumber\\
\hspace{-0.3in}&&\times
\Big[\bar{u}_p\big(\vec{k}_p,\sigma_p\big)\,\gamma^5\,
  u_n\big(\vec{k}_n,\sigma_n\big)\Big] \Big[\bar{u}_e\big(\vec{k}_e,
  \sigma_e\big)\,\gamma_{\mu}(1 - \gamma^5)\,
  v_{\nu}\big(\vec{k}_{\nu}, + \frac{1}{2}\big)\Big]
\end{eqnarray}
where $\Lambda$ is an ultra--violet cut--off.  Thus, to leading order
in the large nucleon mass expansion and to leading order in the
$1/m^2_{\pi}$ expansion \cite{Ivanov2018c} the contribution of a gauge
non--invariant part of the Feynman diagrams in Fig.\,\ref{fig:fig2}d -
Fig.\,\ref{fig:fig2}m does not depend on the electron energy in
complete agreement with Sirlin's analysis of the contribution of
hadronic structure of the nucleon to the radiative corrections of the
neutron lifetime \cite{Sirlin1967,Sirlin1978}. 

The calculation of the gauge invariant amplitude Eq.(\ref{eq:12}) we
carry out in the Feynman gauge at $\xi = 0$ \cite{Sirlin1967} (see
also \cite{Ivanov2017b,Ivanov2013}).  The infrared divergent
contribution, dependent on the electron energy, comes from the Feynman
diagram in Fig.\,\ref{fig:fig2}d. Using the results, obtained in
\cite{Ivanov2013}, for the amplitude of the neutron $\beta^-$--decay,
defined by the Feynman diagrams in Fig.\,\ref{fig:fig2}d -
Fig.\,\ref{fig:fig2}i, we obtain the following expression
\begin{eqnarray}\label{eq:13}
\hspace{-0.3in}&&M(n \to p e^-\bar{\nu}_e)_{\rm
  Fig.\,\ref{fig:fig2}e - Fig.\,\ref{fig:fig2}i} = - G_V\,\Big\{-
\frac{2g_A m_N
  m_e}{m^2_{\pi}}\,\frac{\alpha}{\pi}\Big(\bar{f}_{\beta}(E_e, \mu) +
C^{(\rm div.)}(\Lambda, \mu) + C(E_e)\Big)\Big\}.
\end{eqnarray}
Here $\bar{f}_{\beta}(E_e, \mu)$ is the infrared divergent function of
the electron energy equal to
\begin{eqnarray}\label{eq:14}
\hspace{-0.3in}&&\bar{f}_{\beta}(E_e) = - 2\,{\ell
  n}\frac{\mu}{m_e}\,\Big[\frac{1}{2\beta}\,{\ell n}\Big(\frac{1 +
    \beta}{1 - \beta}\Big) - 1\Big] + \frac{1}{4\beta}\,{\ell
  n}^2\Big(\frac{1 + \beta}{1 - \beta}\Big) + \frac{1}{\beta}\,{\rm
  Li}_2\Big(\frac{2\beta}{1 + \beta}\Big).
\end{eqnarray}
Then, $C^{(\rm div.)}(\Lambda, \mu)$ is the ultra--violet and infrared
divergent expression
\begin{eqnarray}\label{eq:15}
C^{(\rm div.)}(\Lambda, \mu) = 5\,{\ell n}\Big(\frac{\Lambda}{m_N}
\Big) + 6\,{\ell n}\Big(\frac{m_e}{\mu}\Big) + 11\, {\ell
  n}\Big(\frac{m_N} {m_{\pi}} \Big) + 6\, {\ell
  n}\Big(\frac{m_{\pi}}{m_e}\Big) - (4 - \xi) \Big({\ell
    n}\Big(\frac{\Lambda}{m_N}\Big) + {\ell
    n}\Big(\frac{m_N}{\mu}\Big) - \frac{1}{4}\Big)
\end{eqnarray} 
and $C(E_e)$ is a function, dependent on masses of interacting
particles and containing contributions independent of and dependent on
the electron energy. Since it is rather complicated but not important
function from the practical point of view we do not adduce it. The
contribution of the amplitude Eq.(\ref{eq:13}) to the rate of the
neutron $\beta^-$--decay is equal to
\begin{eqnarray}\label{eq:16}
\hspace{-0.3in}&&\lambda^{(\rm h.s.)}_{\beta}(E_0,\mu) =
\frac{\alpha}{\pi}\,(1 + 3
g^2_A)\,\frac{|G_V|^2}{\pi^3}\int^{E_0}_{m_e}dE_e \,\sqrt{E^2_e -
  m^2_e}\,E_e\,F(E_e, Z = 1)\,(E_0 - E_e )^2\,\frac{2 g^2_A}{1 + 3
  g^2_A}\,\frac{ m^2_e}{m^2_{\pi}}\nonumber\\
 \hspace{-0.3in}&&\times \,\Big\{1 + \frac{\alpha}{\pi}\Big(- 2\,{\ell
   n}\frac{\mu}{m_e}\,\Big[\frac{1}{2\beta}\,{\ell n}\Big(\frac{1 +
     \beta}{1 - \beta}\Big) - 1\Big] + \frac{1}{4\beta}\,{\ell
   n}^2\Big(\frac{1 + \beta}{1 - \beta}\Big) + \frac{1}{\beta}\,{\rm
   Li}_2\Big(\frac{2\beta}{1 + \beta}\Big) + C^{(\rm div.)}(\Lambda,
 \mu) + C(E_e)\Big)\Big\}.
\end{eqnarray}
Summing up this contribution with that from the neutron radiative
$\beta^-$--decay \cite{Ivanov2018c}
\begin{eqnarray}\label{eq:17}
\hspace{-0.3in}&&\lambda^{(\rm h.s.)}_{\beta\gamma}(E_0,\mu) =
\frac{\alpha}{\pi}\,(1 + 3
g^2_A)\,\frac{|G_V|^2}{\pi^3}\int^{E_0}_{m_e}dE_e \,\sqrt{E^2_e -
  m^2_e}\,E_e\,F(E_e, Z = 1)\,(E_0 - E_e )^2 \,\frac{2 g^2_A}{1 + 3
  g^2_A}\,\frac{ m^2_e}{m^2_{\pi}}\nonumber\\
 \hspace{-0.3in}&&\times \,\Big\{ - 2{\ell n}\Big(\frac{2 (E_0 -
   E_e)}{\mu}\Big)\Big[\frac{1}{2\beta}{\ell n}\Big(\frac{1 + \beta}{1
     - \beta}\Big) - 1\Big] - 1 - \frac{1}{2\beta}\,{\ell
   n}\Big(\frac{1 + \beta}{1 - \beta}\Big) + \frac{1}{4\beta}\,{\ell
   n}^2\Big(\frac{1 + \beta}{1 - \beta}\Big) + 
 \frac{1}{\beta}\,{\rm Li}_2\Big(\frac{2 \beta}{1 + \beta} \Big)\Big\},
\end{eqnarray}
where $E_0 = (m^2_n - m^2_p + m^2_e)/2m_n = 1.2927\,{\rm MeV}$ is the
end--point energy of the electron--energy spectrum of the neutron
$\beta^-$--decay \cite{Ivanov2013}, we arrive at the contribution of
the hadronic structure of the nucleon to the radiative corrections of
the neutron lifetime
\begin{eqnarray}\label{eq:18}
\hspace{-0.3in}\lambda^{(\rm h.s.)}_n(E_0) &=& \frac{\alpha}{\pi}\,(1
+ 3 g^2_A)\,\frac{|G_V|^2}{\pi^3}\int^{E_0}_{m_e}dE_e \,\sqrt{E^2_e -
  m^2_e}\,E_e\,F(E_e, Z = 1)\,(E_0 - E_e )^2 \nonumber\\
 \hspace{-0.3in}&&\times \,\frac{2 g^2_A}{1 + 3 g^2_A}\,\frac{
   m^2_e}{m^2_{\pi}}\,\Big(1 + \frac{\alpha}{\pi}\,\big(C^{(\rm
   div.)}(\Lambda, \mu) - d_V - d_A\big)\Big)\,\Big(1 +
 \frac{\alpha}{\pi}\, \bar{g}^{(\rm h.s.)}_n(E_e)\Big),
\end{eqnarray}
where $\bar{g}^{(\rm h.s.)}_n(E_e)$ takes the form
\begin{eqnarray}\label{eq:19}
\hspace{-0.3in}\bar{g}^{(\rm h.s.)}_n(E_e) &=& - 2{\ell
  n}\Big(\frac{2 (E_0 - E_e)}{m_e}\Big)\Big[\frac{1}{2\beta}{\ell
    n}\Big(\frac{1 + \beta}{1 - \beta}\Big) - 1\Big] - 1 -
\frac{1}{2\beta}\,{\ell n}\Big(\frac{1 + \beta}{1 - \beta}\Big) +
\frac{1}{2\beta}\,{\ell n}^2\Big(\frac{1 + \beta}{1 - \beta}\Big) +
\frac{2}{\beta}\,{\rm Li}_2\Big(\frac{2 \beta}{1 + \beta}\Big)\nonumber\\
 \hspace{-0.3in}&& +  C(E_e).
\end{eqnarray}
Thus, we have shown that the infrared divergent contributions, induced
by hadronic structure of the nucleon and by virtual photon exchanges
in the neutron $\beta^-$--decay, are cancelled and defined by the
function $\bar{g}^{(\rm h.s.)}_n(E_e)$. The parameters $d_V$ and $d_A$
appear after a formal renormalization of the Fermi weak coupling
constant and the axial coupling constant
\begin{eqnarray}\label{eq:20}
\hspace{-0.3in}G_F\Big(1 + \frac{\alpha}{2\pi}\,d_V\Big) \to G_F
\quad, \quad g_A\Big(1 + \frac{\alpha}{2\pi}\,d_A\Big) \to g_A.
\end{eqnarray}
Since there is no a parameter, which can absorb such a divergent
contribution $1 + (\alpha/\pi)\,\big(C^{(\rm div.)}(\Lambda, \mu) -
d_V - d_A\big)$, so the calculated radiative correction to the neutron
lifetime, induced by hadronic structure of the nucleon and defined by
the function $\bar{g}^{(\rm h.s.)}_n(E_e)$, is not renormalizable and,
correspondingly, is not observable. Following \cite{Ivanov2018b} one
may also assert that such a divergent contribution $C^{(\rm
  div.)}(\Lambda, \mu)$ as well as the contributions of the parameters
$d_V$ and $d_A$ can be hardly cancelled by contributions of other
Feynman diagrams, describing interactions of hadronic structure of the
nucleon to virtual photons, which can be obtained from the Feynman
diagrams in Fig.\,7 and Fig.\,8 in Ref.\,\cite{Ivanov2018b} hooking
photon lines by lines of virtual and real charged particles.

\section{Conclusive discussions}
\label{sec:conclusion}

The aim of this paper has been already formulated in
\cite{Ivanov2018c}. As has been pointed out in \cite{Ivanov2018c}, the
infrared divergent contribution, dependent on the electron energy and
caused by hadronic structure of the nucleon, to the rate of the
neutron radiative $\beta^-$--decay should be cancelled by the
corresponding infrared divergent term, caused by virtual photons
coupled to hadronic structure of the nucleon, in the rate of the
neutron $\beta^-$--decay. Following \cite{Ivanov2018b,Ivanov2018c} we
have performed corresponding calculation within the standard $V - A$
effective theory of weak interactions, QED and the L$\sigma$M,
describing strong low--energy interactions. To the number of Feynman
diagrams, which were calculated by Sirlin \cite{Sirlin1967} with
contributions of strong low--energy interactions given by the axial
coupling constant $g_A$, we have added the Feynman diagrams, caused by
hadronic structure of the nucleon in the form of the one--pion--pole
exchanges and all possible virtual photon exchanges (see
Fig.\,\ref{fig:fig2}). These diagrams are defined by the mesonic part
of the axial--vector hadronic current \cite{Ivanov2018b} and appear
naturally in the standard $V - A$ effective theory of weak
interactions with QED and the L$\sigma$M. Indeed, the Feynman diagrams
in Fig.\,\ref{fig:fig2} can be obtained from the Feynman diagrams in
Fig.\,5 of Ref.\cite{Ivanov2018b} with photon lines hooked by lines of
charged particles. However, unlike the Feynman diagrams in Fig.\,5c -
Fig.\,5f of Ref.\cite{Ivanov2018b}, the contribution of which is
invariant under a gauge transformation of the photon wave function
$\varepsilon^*_{\lambda}(k) \to \varepsilon^*_{\lambda}(k) + c\,k$,
where $c$ is a constant and $\varepsilon^*_{\lambda}(k)$ and $k$ are
the polarization vector and 4--momentum of a real photon obeying the
constraints $k\cdot \varepsilon^*_{\lambda}(k) = 0$ and $k^2 = 0$, the
contribution of the Feynman diagrams in Fig.\,\ref{fig:fig2}d -
Fig.\,\ref{fig:fig2}f is not invariant under a gauge transformation of
the photon propagator $D_{\alpha\beta}(k) \to D_{\alpha\beta}(k) +
c(k^2)\,k_{\alpha}k_{\beta}$, where $c(k^2)$ is an arbitrary function
of $k^2$ and $k$ is a 4--momentum of a virtual photon $k^2 \neq 0$.

We have found that a gauge non--invariant part of the Feynman diagrams
in Fig.\,\ref{fig:fig2}d - Fig.\,\ref{fig:fig2}f, describing the
contribution of hadronic structure of the nucleon to the radiative
corrections, is ultra--violet and infrared divergent but does not
depend on the electron energy. This agrees well with Sirlin's analysis
of hadronic structure of the nucleon \cite{Sirlin1967,Sirlin1978}. We
have got the required infrared divergent contribution to the rate of
the neutron $\beta^-$--decay, which cancels infrared divergence to the
neutron lifetime from the infrared divergent contribution of hadronic
structure of the nucleon to the rate of the neutron radiative
$\beta^-$--decay obtained in \cite{Ivanov2018c}. At this level we may
confess that our doubts, concerning impossibility of such a
cancellation within the standard $V - A$ effective theory of weak
interactions with QED and the L$\sigma$M, have not come true.

Nevertheless, the ultra--violet and infrared divergent contributions
independent of the electron energy $1 + (\alpha/\pi)\,\big(C^{(\rm
  div.)}(\Lambda, \mu) - d_V - d_A\big)$, accompanying the calculation
of the required infrared divergent expression dependent on the
electron energy, cannot be removed by even a formal renormalization
because of the absence of any parameter similar to the Fermi weak
coupling constant or the axial coupling constant, which could absorb
such a divergent term. This testifies unrenormalizability and,
correspondingly, unobservability of the obtained radiative corrections
of order $O(\alpha/\pi)$ to the neutron lifetime, induced by hadronic
structure of the nucleon. In spit of the fact that the relative
contribution of these corrections, described by the function
$\bar{g}^{(\rm h.s.)}_n(E_e)$ (see Eq.(\ref{eq:19})), to the neutron
lifetime is of order $(\alpha/ \pi)\,10^{-5} \sim 10^{-9}$, the
problem of unrenormalizability of contributions of hadron--photon
virtual interactions within the standard $V - A$ effective theory of
weak interactions with QED and L$\sigma$M makes such a theory
inapplicable to the calculation higher oder radiative corrections such
as $O(\alpha^2/\pi^2)$ and also corrections of order $O(\alpha
E_e/m_N)$ \cite{Ivanov2018b,Ivanov2018c}.  The contributions of other
Feynman diagrams with virtual photon exchanges, which can be obtained
from the Feynman diagrams in Fig.\,7 and Fig.\,8 of
Ref.\cite{Ivanov2018b} with photon lines hooked by lines of charged
virtual and real particles, can hardly solve the problem of
renormalizability of the radiative corrections of order
$O(\alpha/\pi)$ to neutron lifetime, caused by hadronic structure of
the nucleon. As has been pointed out in \cite{Ivanov2018b} the problem
of unrenormalizability and as well as of non--invariance under gauge
transformations is related to the use of the effective $V - A$ theory
of weak interaction, where the $V - A$ vertex of hadron--lepton
current--current interactions is not the vertex of the combined
quantum field theory including QED and the L$\sigma$M.

Thus, as has been argued in \cite{Ivanov2018b,Ivanov2018c} a
consistent gauge invariant and renormalizable analysis of radiative
corrections, induced by hadronic structure of the nucleon, can be
performed only in a fully gauge invariant and renormalizable quantum
field theory including the Standard Electroweak Model (SEM) and a
renormalizable theory of strong low--energy interactions, e.g. the
L$\sigma$M. According to \cite{Ivanov2017b}, the use of such a
combined quantum field theory should be of great importance for a
consistent analysis of Standard Model corrections of order $10^{-5}$
\cite{Ivanov2017b}, which should provide a robust theoretical
background for searches of contributions of interactions beyond the
Standard Model, caused by first and second class hadronic currents of
hadron--lepton current--current interactions \cite{Ivanov2018a} (see
also \cite{Gardner2001,Gardner2013a}).

\section{Acknowledgements}

We thank Hartmut Abele for discussions stimulating the work under this
paper as a step towards the analysis of the Standard Model corrections
of order $10^{-5}$ \cite{Ivanov2018b,Ivanov2017b}.  The work of
A. N. Ivanov was supported by the Austrian ``Fonds zur F\"orderung der
Wissenschaftlichen Forschung'' (FWF) under contracts P26781-N20 and
P26636-N20 and ``Deutsche F\"orderungsgemeinschaft'' (DFG) AB
128/5-2. The work of R. H\"ollwieser was supported by the Deutsche
Forschungsgemeinschaft in the SFB/TR 55. The work of M. Wellenzohn was
supported by the MA 23 (FH-Call 16) under the project ``Photonik -
Stiftungsprofessur f\"ur Lehre''.

\end{document}